\documentclass[prd, twocolumn,nofootinbib,superscriptaddress]{revtex4}

\usepackage{subfigure}
\usepackage{diagbox}
\usepackage{multirow}  
\usepackage{epsfig,latexsym,cancel,amssymb,amsmath,verbatim,mathrsfs}
\usepackage{color,xcolor,graphicx}

\usepackage{hyperref}
\hypersetup{
	colorlinks,
	linkcolor={red!70!black},
	citecolor={blue!80!black},
	urlcolor={blue!80!black},
	bookmarksopen=true
}
\newcommand{\beq}{\begin{equation}}
\newcommand{\eeq}{\end{equation}}
\newcommand{\bea}{\begin{eqnarray}}
\newcommand{\eea}{\end{eqnarray}}

\newcommand{\nn}{\nonumber}

\definecolor{darkgreen}{HTML}{228B22}

\allowdisplaybreaks[4]

\begin{document}

\title{Probing Magnetic Moment Operators in $H \gamma$ Production and $H \to \tau^+ \tau^- \gamma$ Rare Decay}
\author{Qing-Hong Cao}
\email{qinghongcao@pku.edu.cn}
\affiliation{School of Physics and State Key Laboratory 
of Nuclear Physics and Technology, Peking University, Beijing 100871, China}
\affiliation{Collaborative Innovation Center of Quantum Matter, Beijing 100871, China}
\affiliation{Center for High Energy Physics, Peking University, Beijing 100871, China}

\author{Hao-Ran Jiang}
\email{h.r.jiang@pku.edu.cn}
\affiliation{School of Physics and State Key Laboratory 
of Nuclear Physics and Technology, Peking University, Beijing 100871, China}
\author{Bin Li}
\email{libin@pku.edu.cn}
\affiliation{School of Physics and State Key Laboratory 
of Nuclear Physics and Technology, Peking University, Beijing 100871, China}

\author{Yandong Liu}
\email{ydliu@bnu.edu.cn}
\affiliation{Key Laboratory of Beam Technology of Ministry of Education, College of Nuclear Science and Technology, Beijing Normal University, Beijing 100875, China}
\affiliation{Beijing Radiation Center, Beijing 100875, China}

\author{Guojin Zeng}
\email{guojintseng@pku.edu.cn}
\affiliation{School of Physics and State Key Laboratory 
of Nuclear Physics and Technology, Peking University, Beijing 100871, China}

\begin{abstract}
The magnetic moment ($a_\gamma$) and weak magnetic moment ($a_W$) of charged leptons and quarks are sensitive to quantum effects of new physics heavy resonances. In effective field theory $a_\gamma$ and $a_W$ are induced by two independent operators, therefore, one has to measure both the $a_\gamma$ and $a_W$ to shed lights on new physics. The $a_W$'s of the SM fermions are measured at the LEP. In this work, we analyze the contributions from magnetic and weak magnetic moment operators in the processes of $pp\to H \gamma$ and $gg\to H \to \tau^+ \tau^- \gamma$ at the High-Luminosity Large Hadron Collider. We demonstrate that the two processes could cover most of the parameter space that cannot be probed at the LEP.
\end{abstract}

\maketitle

\noindent{\bf 1. Introduction.}
\label{section1}

\begin{figure}
\includegraphics[scale=0.65]{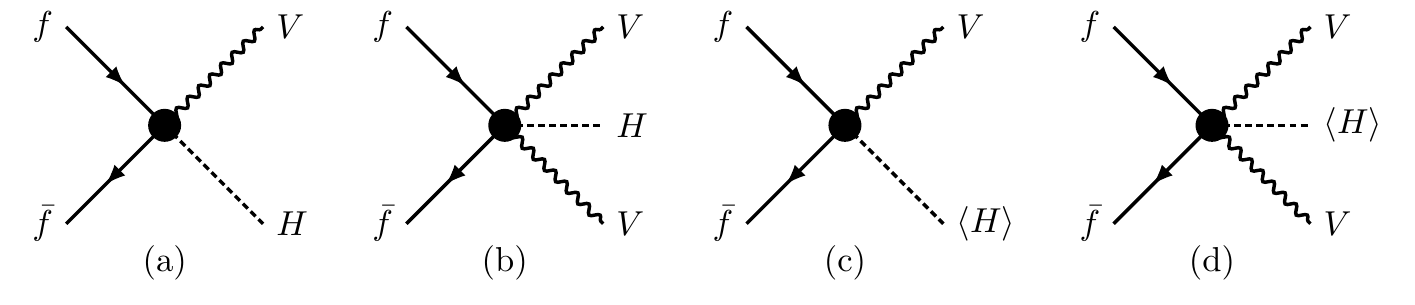}
\caption{Feynman diagrams of the dimension-6 operator before and after the symmetry breaking. } 
\label{fig:op}
\end{figure}

Searching for new physics (NP) beyond the Standard Model (SM) is the key mission of particle physics. Although no heavy resonances have been discovered at the Large Hadron Collider (LHC), one can probe the quantum effects of those heavy resonances through the measurement of magnetic moment  ($a_\gamma$) and weak magnetic moment ($a_W$) of the SM fermions~\cite{Miller:2007kk,Lindner:2016bgg,Stockinger:2006zn,Jegerlehner:2009ry}. When NP resonances are too heavy to be directly probed at the current colliders, one could describe the unknown NP effects through high-dimensional operators constructed with the SM fields at the NP scale $\Lambda$, obeying the well-established gauge structure of the SM, i.e. $SU(2)_{W}\otimes U(1)_Y$. The Lagrangian of effective field theory (EFT) is 
\begin{equation}
\mathcal{L}_{\rm EFT}=\mathcal{L}_{SM}+\frac{1}{\Lambda^{2}}\sum_{i}\left(C_{i}{O}_{i}+h.c.\right)+\mathcal{O}\left(1/\Lambda^{3}\right),\label{eq:eft}\end{equation}
where $C_{i}$'s are the Wilson coefficients. 
In the Warsaw basis the dimension-6 operators $O_{fW}$'s and $O_{fB}$'s that generate $a_\gamma$ and $a_W$ are given by~\cite{WarsawBasis,SMEFTFR:2017zog}
\begin{align}
O_{e_iW}&=(\bar{L}_i \sigma^{\mu\nu} \tau^I e_i)  \phi W^I_{\mu\nu},\nonumber\\
O_{e_iB}&=(\bar{L}_i \sigma^{\mu\nu}e_i) \phi B_{\mu\nu},\nonumber\\
O_{u_iW}&=(\bar{Q}_i \sigma^{\mu\nu} \tau^I u_i) \tilde\phi W^I_{\mu\nu},\nonumber\\
O_{u_iB}&=(\bar{Q}_i \sigma^{\mu\nu}u_i) \tilde\phi B_{\mu\nu},\nn\\
O_{d_iW}&=(\bar{Q}_i \sigma^{\mu\nu}\tau^I d_i) \phi W^I_{\mu\nu},\nonumber\\ 
O_{d_iB}&=(\bar{Q}_i \sigma^{\mu\nu} d_i) \phi B_{\mu\nu},
\end{align}
where $L_i$ and $Q_i$ denotes the left-handed weak doublet of the $i$-th generation in the SM while $e_i (u_i, d_i)$ the right-handed weak singlet of charged leptons (up-type and down-type quarks), respectively.  
Figure~\ref{fig:op} shows Feynman diagrams of the dimension-6 operators; see (a) and  (b). After spontaneously symmetry breaking the operators yield $f\bar{f}V$ and $f\bar{f}VV$ anomalous couplings; see (c) and (d). The $f\bar{f}V$ anomalous couplings give rise to the magnetic moment ($a_\gamma^f$) and the weak magnetic moment ($a_W^f$) of the fermion $f$ as follows:
\begin{align}
\delta a_\gamma^{f=u,c,t} = -2 \sqrt{2} \frac{m_f v}{\Lambda^2} \frac{1}{e Q_f} (c_W~C_{fB}+s_W~C_{fW}) ,\nonumber\\ 
\delta a_W^{f=u,c,t} = +2 \sqrt{2} \frac{m_f v}{\Lambda^2} \frac{1}{e Q_f} (s_W~C_{fB}-c_W~C_{fW}) ,\nonumber\\
\delta a_\gamma^{f=\ell,d,s,b}  = -2 \sqrt{2} \frac{m_f v}{\Lambda^2} \frac{1}{e Q_f} (c_W~C_{fB}-s_W~C_{fW}) ,\nonumber\\
\delta a_W^{f=\ell,d,s,b} =+2 \sqrt{2} \frac{m_f v}{\Lambda^2} \frac{1}{e Q_f} (s_W~C_{fB}+c_W~C_{fW}),
\label{eq:MM}
\end{align}
where $Q_f$ and $m_f$ denotes the charge and mass of the fermion $f$, $v=246~{\rm GeV}$ is the vacuum expectation value of the Higgs doublet after symmetry breaking, $s_W$ and $c_W$ denotes sine and cosine of the Weinberg angle, respectively. As the $a_\gamma^f$ and $a_W^f$ are orthogonal in the parameter space of $C_{fB}$ and $C_{fW}$, one has to measure both the $a_\gamma^f$ and $a_W^f$ to probe the NP effects.

The magnetic moments of up-quarks and down-quarks $a_\gamma^{u,d}$ and $a_W^{u,d}$ (corresponding to the operators $O_{uW/dW}$ and $O_{uB/dB}$) are tightly constrained through Drell-Yan processes, $VV$ pair production and $VH$ associated production at the LHC~\cite{Almeida:2019jqy,Aad:2020jym}. The operators $O_{tW}$ and $O_{tB}$ of top quarks could be examined in single-top productions or top-quark decays~\cite{Zhang:2010dr,Cao:2021wcc,Agashe:2014kda,bEDM,Goldouzian:2020ekx}. The $a_W^{s,c,b}$ is bounded by the precise measurements at the LEP~\cite{Abdallah:2003xd,Rizzo:1994qz,Alcaraz:2006mx,Escribano:1993xr}, which yield 
\begin{align}
& \Big| s_W C_{s(b) B} + c_W C_{s(b) W}\Big|\left(\frac{1~ \text{TeV}}{\Lambda}\right)^2 \leq 4.2,\nn\\
& \Big| s_W C_{c B} - c_W C_{c W}\Big| \left(\frac{1~ \text{TeV}}{\Lambda}\right)^2\leq  4.2~.
\end{align}
In this work we show that the $a_\gamma^{s,c,b}$ induced by the two operators can be tested in the process of $pp\to H \gamma$ at the LHC with an integrated luminosity of $3000~{\rm fb}^{-1}$ (HL-LHC).  

The magnetic moments of electrons and muons are severely constrained by the $Z$-boson width measurement at the LEP~\cite{Escribano:1993xr} and the measurements of magnetic moment~\cite{Parker_2018, Hanneke_2008, Hanneke_2011, PhysRevD.73.013003,Bennett_2006, Davier_2017, Keshavarzi_2018,Abi:2021gix}; therefore, we do not consider the electron and muon in this work. The LEP constraint on the $a_W^\tau$ of the $\tau$-lepton reads as~\cite{Abdallah:2003xd,Rizzo:1994qz,Alcaraz:2006mx,Escribano:1993xr,Heister:2002ik,GonzalezSprinberg:2000mk}
\begin{equation}
\Big| s_W C_{\tau B} + c_W C_{\tau W}\Big|\left(\frac{1~ \text{TeV}}{\Lambda}\right)^2 \leq 0.28
 ~,
\end{equation}
while the constraint on the $a_\gamma^\tau$ as
\begin{equation}
 -12.8  \leq \big(c_W C_{\tau B} - s_W C_{\tau W}\big) \left(\frac{1~ \text{TeV}}{\Lambda}\right)^2 \leq 3.2
 ~.
\end{equation}
We demonstrate that the $a_\gamma^\tau$ can be measured in the process of $gg \to H \to \tau^+ \tau^- \gamma$ at the HL-LHC.

~\\
\noindent{\bf 2. The $H\gamma$ associated production.}
\label{hadronhgamma}

\begin{figure}
\includegraphics[scale=0.75]{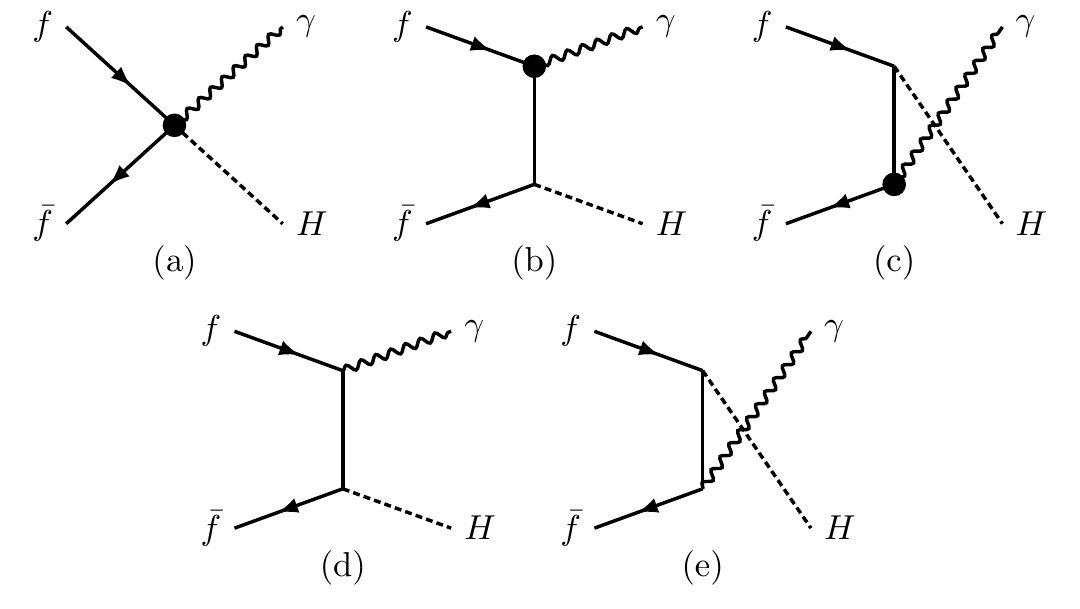}
\caption{Feynman diagrams of the $H \gamma$ production. } 
\label{fig:FeynmanDiagram}
\end{figure}

In this section we examine the effects of magnetic-moment operators in the $pp\to H \gamma$ process at the HL-LHC, which has been studied extensively in the literature~\cite{CMS:2018ars,Aaboud:2018fgi,Aad:2020ylk,Shi:2018lqf,Khanpour:2017inb}. 
We consider one flavor at a time throughout this work. Figure~\ref{fig:FeynmanDiagram}(a) and (b, c) display the Feynman diagrams of the $H\gamma$ production induced by the operators $O_{fB}$ and $O_{fW}$. 
The SM process is shown in Fig.~\ref{fig:FeynmanDiagram}(d,~e). There are non-zero interference effects between the operator-induced diagrams and the SM diagrams, therefore, we also treat the interference effect as the signal. However, as explained below, only the first diagram contributes after applying  hard kinematic cuts and the interference effects are negligible. 

In our simulation the Higgs bosons are required to decay into a pair of bottom quarks, the predominant decay mode of the Higgs boson.  The event topology of the signal process is two bottom quarks plus a photon. The SM backgrounds are 
\begin{eqnarray}
&&pp\to \gamma +\text{jets},\nonumber\\
&&pp\to t \bar{t} \gamma,\nonumber\\
&&pp\to Z \gamma ,\nonumber\\
&&pp\to t\gamma/\bar{t}\gamma  +\text{jets},
\end{eqnarray}
with the $Z$ boson and top quark hadronic decay. It is noted that background $\gamma+\text{jets}$ consists of $\gamma+b\bar{b}$, $\gamma+c\bar{c}$, $\gamma~+$ light flavor jets and etc. 

We generate signal and background events utilizing MadEvent~\cite{madgraph}, and then pass those to Pythia~\cite{pythia} and Delphes~\cite{delphes} for parton showering, hadronization and collider simulation. In order to avoid collinear and soft divergences in the process of $pp\to \gamma+\text{jets}$ we apply the kinematic cuts at the generator level as follows: 
\begin{align}
&p_\text{T}^\gamma\geq 30~\mathrm{GeV},  ~&&|\eta_{\gamma}|\leq 2.5,\nn\\
&p_\text{T}^j\geq 30~\mathrm{GeV}, ~&&|\eta_{j}|\leq 2.5, \nn\\
&\Delta R_{jj}>0.4,~&&\Delta R_{j\gamma}> 0.4,
\label{eq:cut1}
\end{align}
where $p_{\rm T}^{\gamma/j}$ and $\eta^{\gamma/j}$ is the transverse momentum and pseudo-rapidity of $\gamma$ and jet, respectively, and $\Delta R_{ij}\equiv  \sqrt{(\phi_{i} - \phi_{j})^2 + (\eta_{i}-\eta_{j})^2}$ is the angular distance between the objects $i$ and $j$ in the azimuthal angle ($\phi$)-pseudurapidity ($\eta$) plane. At the detector level two $b$-jets are demanded in the final state to suppress the SM backgrounds. In the simulation we utilize $b$-tagging technology~\cite{Chatrchyan:2012jua,Aad:2015ydr} to distinguish the jet flavor. 
The $b$-tagging efficiency is chosen as $70\%$, the mistagging rate of $c$-quarks and light flavor quarks is $10\%$ and $1\%$, respectively. We require at least one photon in the final state, i.e., 
\begin{equation}
\label{eq:cut1prime}
n^{b-\rm{jet}} = 2,~n^\gamma \geq 1.
\end{equation}

\begin{figure}
\centering
\includegraphics[scale=.7]{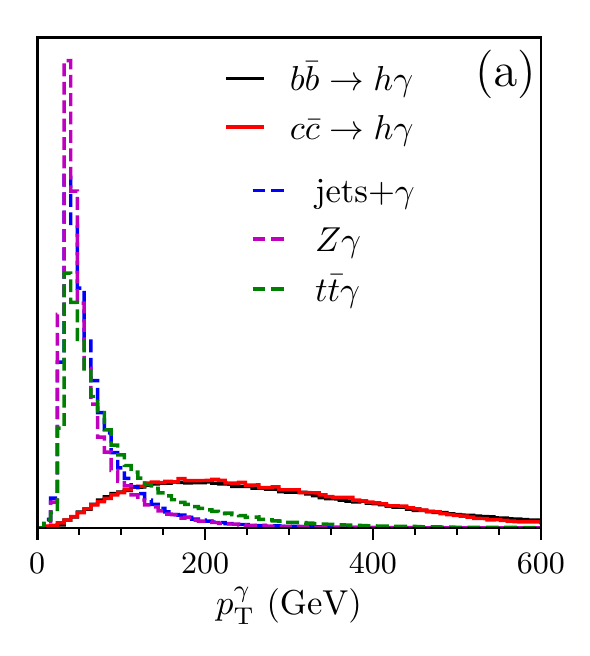}
\includegraphics[scale=.7]{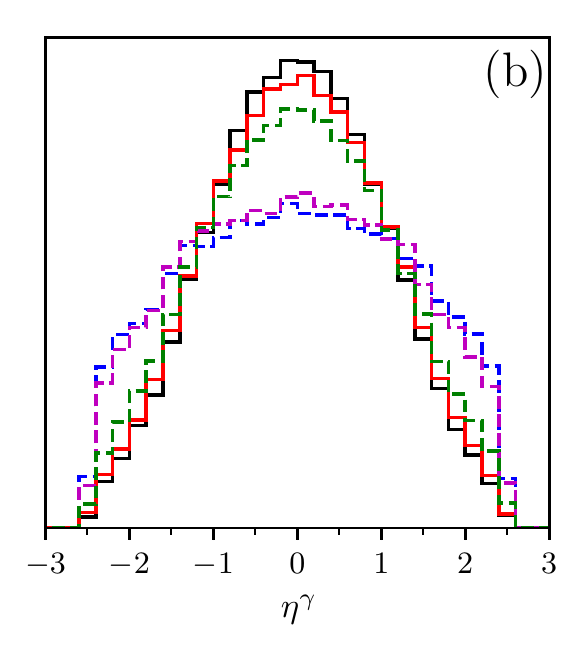}
\includegraphics[scale=.7]{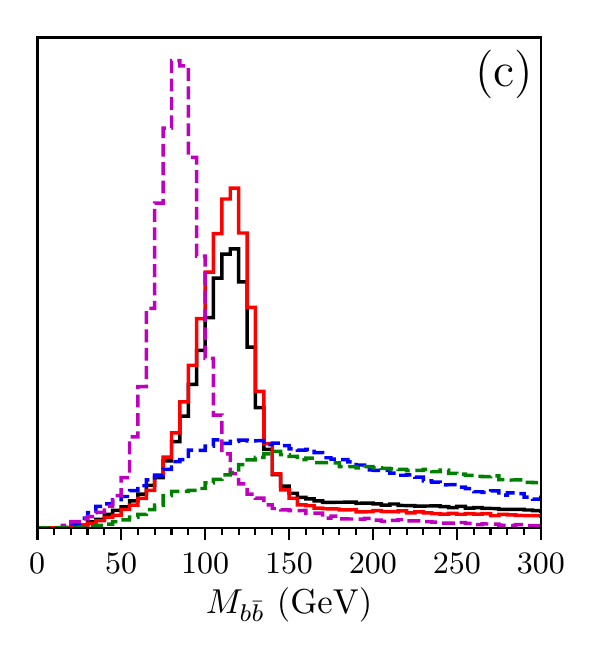}
\includegraphics[scale=.7]{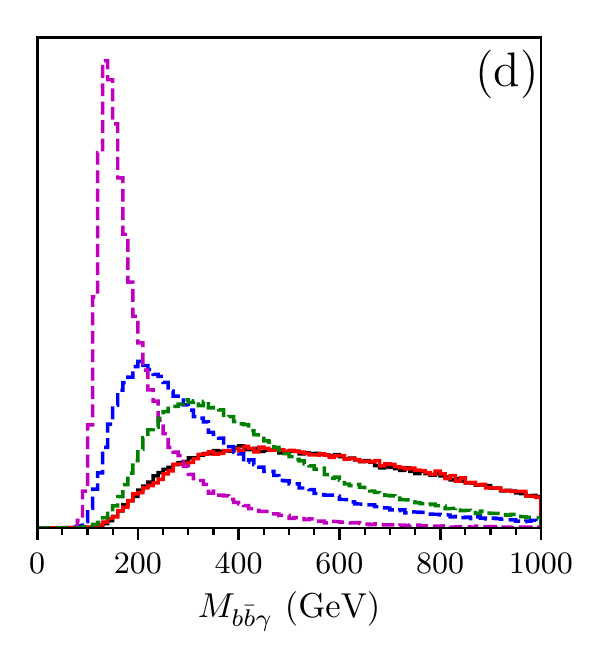}
\caption{The normalized distributions of $P_{\text{T}}^\gamma$ (a), $\eta^\gamma$ (b), $M_{b\bar b}$ (c) and $M_{b\bar b\gamma}$ (d). The black (red) solid curve denotes the distributions of the signal events induced by the operator $O_{bB}$ ($O_{cB}$), respectively. The operator $O_{bW}$ ($O_{cW}$) yields exactly the same normalized distributions as the operator $O_{bB}$ ($O_{cB}$), respectively. The dashed curves label the SM backgrounds from the $\gamma+{\rm jets}$ (blue), $Z\gamma$ (magenta) and $t \bar{t}\gamma$ (green), respectively. } 
\label{Distributionhgamma}
\end{figure}

Figure~\ref{Distributionhgamma} displays the normalized distributions of $p_{\text T}^\gamma$ (a), $\eta^\gamma$ (b), $M_{b\bar{b}}$ (c) and $M_{b\bar{b}\gamma}$ (d) after imposing the \underline{\it basic cuts} given in Eq.~\ref{eq:cut1} and Eq.~\ref{eq:cut1prime}. For demonstration, we plot the distributions of the signal events induced by the operator $O_{cB}$ and $O_{bB}$, respectively. It shows that the photons in the signal events exhibit a hard $p_{\text T}$ and mainly appear in the central region of the detector; see the black-solid and the red-solid curves. The reason can be understood as follows. In the signal events the two fermions in the initial state are in the different chirality states and thus are in the $s$-wave state. In order to respect the angular momentum conservation, the particles in the final state are in the $p$-wave state such that the matrix element is proportional to $p_{\rm T}^\gamma$. As a result, the matrix element of the signal process is proportional to $\sin\theta$ where $\theta$ is the polar angle of the photon with respect to the beam line in the center of mass frame. On the other hand, the photons in the background mainly arise from the QED radiation and tend to be soft. 

For illustration, we present the leading contributions of the squared matrix elements of the $f\bar{f}\to H\gamma$ process as follows: 
\begin{align}
&|M_{a}|^2= \frac{4N_c c^2_W  (s-m_H^2)^2}{\Lambda^4}\sin^2\theta, \nn \\
&|M_{b}+M_{c}|^2=\frac{48N_c c^2_W  m_f^2s}{\Lambda^4},\nn \\
&|M_{d}+M_{e}|^2\nn\\
=&\frac{8 N_c Q_f^2 e^2 m_f^2  \big[ 3 s^2- 2 s m_H^2- \cos\theta (s-m_H)^2 +3 m_H^4  \big]  }{  v^2 (s-m_H)^2 \sin^2\theta},\nn \\
&{\rm Re}(M^\dagger_{a}(M_{d}+M_{e}))\nn\\
=&\frac{\sqrt{2} N_c Q_f e c_W  m_f \big[3 s  + \cos\theta(s-m_H^2)  -7 m_H^2  \big] }{\Lambda^2 v}, \label{EQ:sme}
\end{align}
where $N_c=3$ is the color factor, the subscript of the matrix element denotes the corresponding Feynman diagram in Fig.~\ref{fig:FeynmanDiagram}. In the region of large colliding energy, only $|M_{a}|^2$ contributes while the others contributions are negligible. Indeed, the $|M_{a}|^2$ is proportional to $\sin^2\theta$. 

Taking advantage of the hard photon in the signal events, we impose a hard $p_T$ cut on the photon with the largest $p_T$ as following: 
\begin{equation}
p_\text{T}^\gamma\geq 300~\mathrm{GeV},
\label{eq:pta}
\end{equation}
to suppress the SM backgrounds. 

Figure~\ref{Distributionhgamma}(c) displays the normalized distributions of the invariant mass of two $b$-jets ($M_{b \bar{b} }$). The two $b$-jets in the signal event originate from the Higgs boson decay, therefore, their invariant mass is around $m_H$; see the black-solid and the red-solid curves. Similarly, there is a peak around $m_Z$ in the $Z \gamma$ backgrounds. The two $b$-jets in other SM backgrounds are not from a resonance decay and yield a flat $M_{b \bar{b} }$ distribution. We impose a mass window cut on the two $b$-jets,
\begin{equation}
\left|M_{b \bar{b} }-m_H\right| \leq 15~\mathrm{GeV},
\label{eq:mbb}
\end{equation}
to suppress the SM backgrounds.

Figure~\ref{Distributionhgamma}(d) displays the normalized distributions of the invariant mass of two $b$-jets and a photon ($M_{ b \bar{b} \gamma}$). The signal events tend to have a large invariant mass while the background events prefer the small invariant mass region.  In order to suppress the SM backgrounds, we further impose a hard cut on $M_{ b \bar{b} \gamma}$ as following:
 \begin{equation}
M_{ b \bar{b} \gamma} \geq 700~\mathrm{GeV}.
\label{eq:mbba}
\end{equation}

Table~\ref{hgamma} shows the numbers of the signal events and the background events after the basic cuts and the \underline{\it optimal cuts} (i.e., the $p_{\text T}^\gamma$, $M_{b\bar{b}}$ and $M_{b\bar{b}\gamma}$ cuts) at the HL-LHC. Note that the NP scale $\Lambda$ is set to be $1~{\rm TeV}$. 
The major SM background comes from the $\gamma+{\rm jets}$ process in which the $\gamma b\bar{b}$ channel dominates. As both the $O_{fB}$ and $O_{fW}$ operators contribute to the signal process through the same $f\bar{f}VH$ vertex, they generate the same differential distributions and therefore have the same cut efficiencies. The $O_{fW}$ and $O_{fB}$ operators differ in the cross section by a total factor $\tan^2\theta_W$.

\begin{table}
  \caption{The number of the signal and background events at the HL-LHC for $\Lambda = 1~{\rm TeV}$.}
  \label{hgamma}
  \centering
\vspace{0.3cm}
\begin{tabular}{c|c|c}
\hline
Signal processes  & Basic cuts & Optimal cuts\\\hline
$C_{bB}=1$, $C_{bW}=0$ &   $9.38\times 10^{3}$  & $815$\\ \hline
$C_{bB}=0$, $C_{bW}=1$ &     $2.82\times 10^{3}$      & $245$\\
\hline
$C_{cB}=1$, $C_{cW}=0$  &     $2.78\times 10^{4}$     & $3.17\times 10^{3}$\\ \hline
$C_{cB}=0$, $C_{cW}=1$ &     $8.37\times 10^{3}$    & $953$\\
\hline
$C_{sB}=1$, $C_{sW}=0$  &    $3.86\times 10^{4}$    & $4.79\times 10^{3}$\\ \hline
$C_{sB}=0$, $C_{sW}=1$ &     $1.16\times 10^{4}$    & $1.44\times 10^{3}$\\
\hline
\hline
Background processes  & Basic Cuts & Optimal cuts\\\hline
$\gamma$+jets  & $3.32 \times 10^{8}$  &$9.90\times 10^{4}$\\ \hline
$Z\gamma$&     $5.80 \times 10^{5}$    &$828$\\ \hline
$t\bar{t}\gamma$ & $4.85 \times 10^{5}$ &$1.25\times 10^{3}$\\ \hline
$t\gamma/\bar{t}\gamma  +\text{jets}$ &   $2.40 \times 10^{5}$      &$106$\\ \hline
\end{tabular}
\end{table}

\begin{figure*}
\centering
\includegraphics[scale=.57]{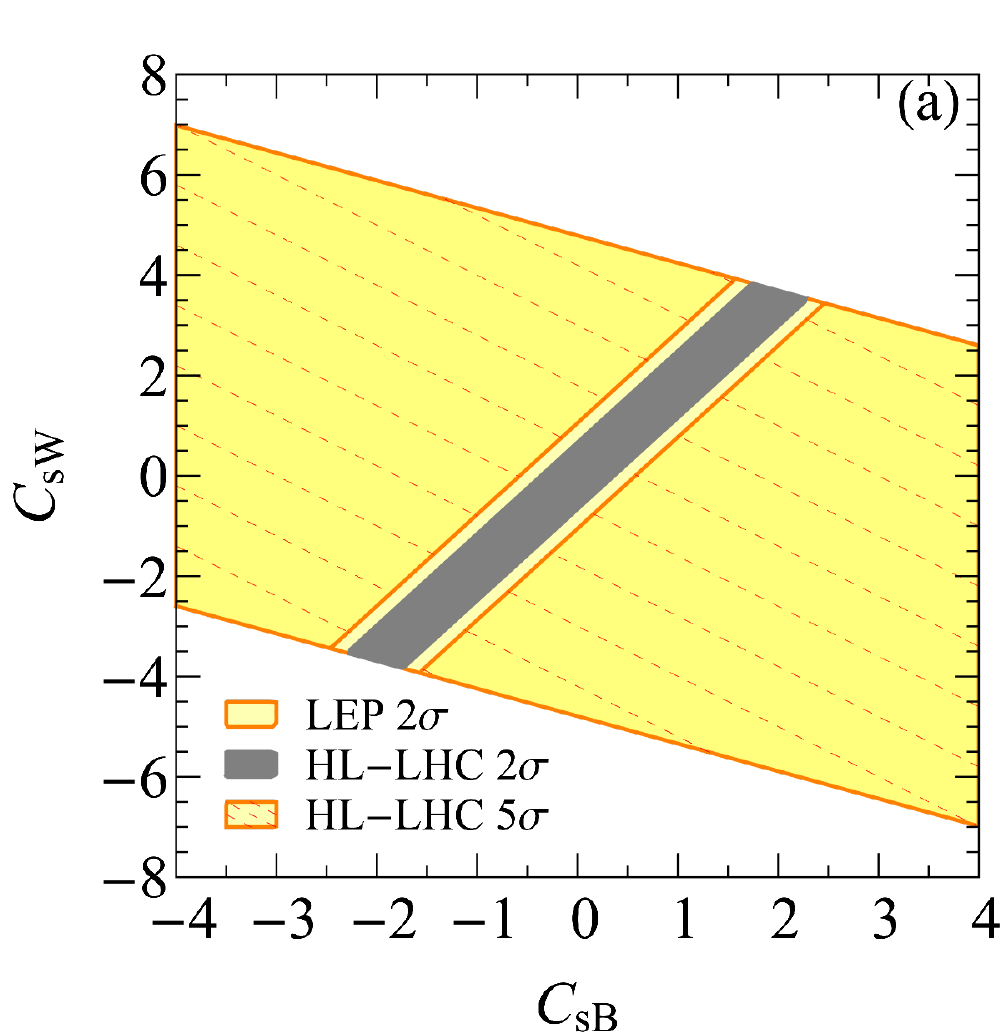}
\includegraphics[scale=.57]{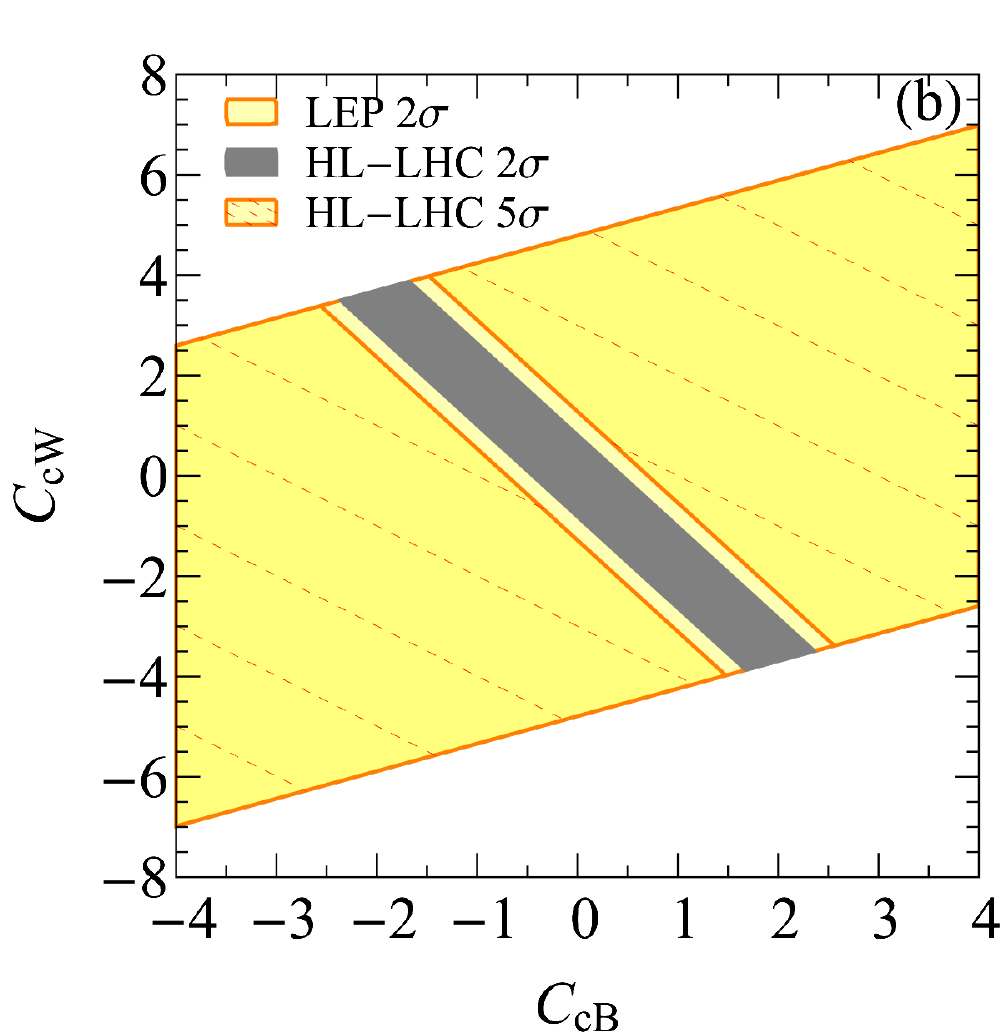}
\includegraphics[scale=.57]{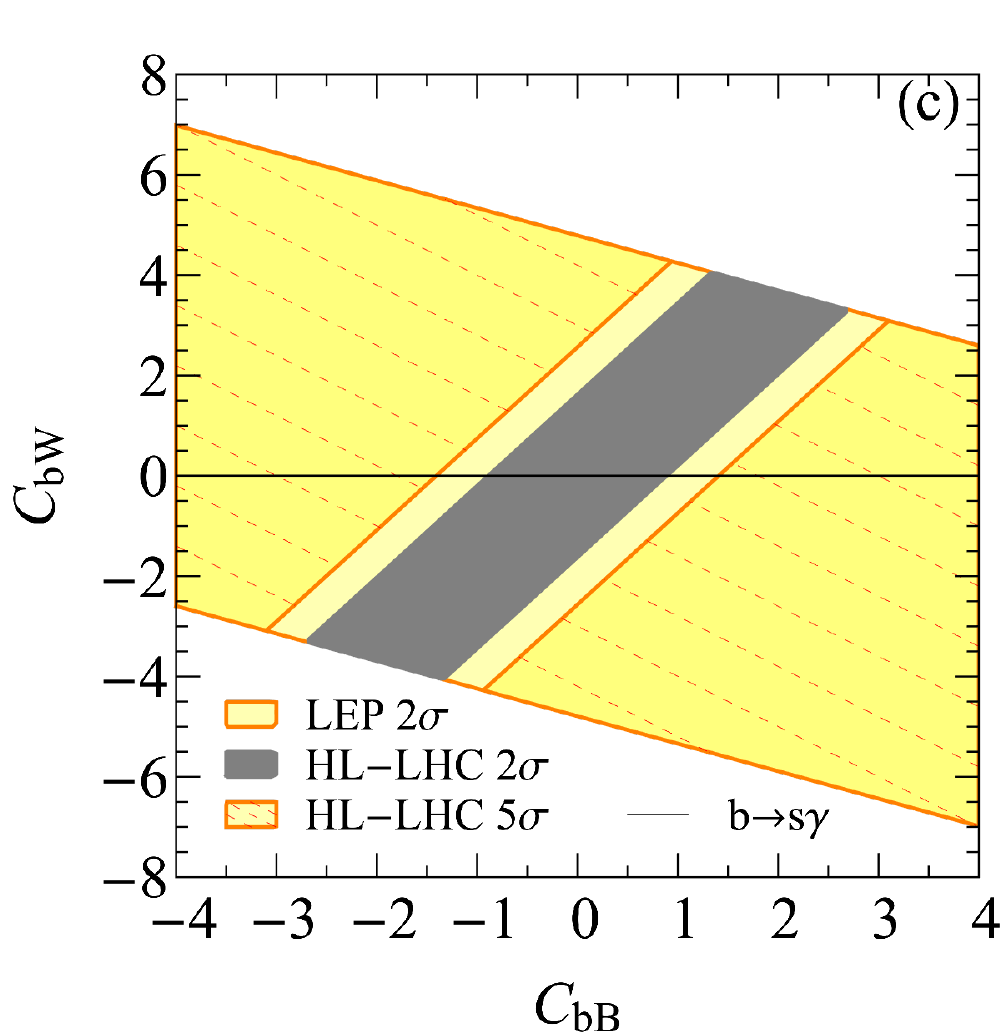}
\caption{The yellow-meshed bands denote the $5\sigma$ discovery region of $C_{fB}$ and $C_{fW}$ for the $s$-quark (a), the $c$-quark (b) and $b$-quark (c) in the $H\gamma$ production with $\Lambda=1~{\rm TeV}$ at the HL-LHC, respectively, while the yellow bands denote the allowed regions by the $2\sigma$ bounds at the LEP. The gray band denotes the allowed region at the $2\sigma$ significance if no NP effects are observed in the $H\gamma$ production.  The black line denotes the $2\sigma$ constraint of the $b\to s\gamma$ measurement. }
\label{cqw_cqb}
\end{figure*}

Equipped with the optimal cuts shown above, we vary the Wilson coefficients to obtain a 5 standard deviations ($\sigma$) statistical significance using 
\beq
\sqrt{-2\left[(n_b + n_s) \log\frac{n_b}{n_s+n_b}+n_s\right]}=5,
\label{eq:dis}
\eeq
where $n_b$ and $n_s$ represents the numbers of the signal and background events, respectively.
The number of the signal events in Table~\ref{hgamma} is calculated with the choice of  $C_{fW}=1$ or $C_{fB}=1$ ($f=c,s,b$), and $\Lambda=1~{\rm TeV}$.  Denote the number of the signal events in the last column of Table~\ref{hgamma} after all the cuts as $n_s^0$. The  $n_s^f$  for a general choice of $C_{fW}$, $O_{fB}$ and $\Lambda$ can be obtained as follows:
\begin{align}
n_s^s = n_s^0\Big|_{C_{sB}=1}\times\frac{(c_W C_{sB} - s_W C_{sW})^2}{c_W^2} \left(\frac{1~ \text{TeV}}{\Lambda}\right)^4 , \nn \\
n_s^c =  n_s^0\Big|_{C_{cB}=1}\times \frac{(c_W C_{cB} + s_W C_{cW})^2}{c_W^2} \left(\frac{1~ \text{TeV}}{\Lambda}\right)^4 , \nn \\ 
n_s^b = n_s^0\Big|_{C_{bB}=1}\times\frac{(c_W C_{bB} - s_W C_{bW})^2}{c_W^2} \left(\frac{1~ \text{TeV}}{\Lambda}\right)^4.
\end{align}
Using Eq.~\ref{eq:dis}, we obtain that a $5\sigma$ discovery significance requires 
\begin{align}
&\Big|c_W C_{sB} - s_W C_{sW}\Big|\left(\frac{1~ \text{TeV}}{\Lambda}\right)^2\geq 0.51,\nonumber\\ 
&\Big|c_W C_{cB} + s_W C_{cW}\Big|\left(\frac{1~ \text{TeV}}{\Lambda}\right)^2\geq 0.62,\nonumber\\ 
&\Big|c_W C_{bB} - s_W C_{bW}\Big|\left(\frac{1~ \text{TeV}}{\Lambda}\right)^2\geq 1.23~.
\end{align}

Figure~\ref{cqw_cqb} displays the $5\sigma$ discovery region in the plane of $C_{f W}$ and $C_{f B}$ for the strange-quark (a), the charm-quark (c) and the bottom-quark (c) in the $H\gamma$ production at the HL-LHC (yellow-meshed band). The yellow band denotes the allowed $2\sigma$ parameter space by the $Z$-width measurement at the LEP. We consider one flavor at a time.  The HL-LHC has a great potential of probing the $O_{s(c,b)B}$ and $O_{s(c,b)W}$ in comparison with the LEP. The operator $O_{bW}$ is  highly constrained by the $b\to s\gamma$ measurement~\cite{Agashe:2014kda,bEDM}, i.e., 
\begin{equation}
-0.008 \leq C_{bW}\leq 0.011~.
\end{equation}
We plot the constraint of $b\to s\gamma$ on the $O_{bW}$ at the $95\%$ confidence level in Fig.~\ref{cqw_cqb}(c); see the black line. Obviously, the constraints from the $b\to s\gamma$ measurement is much more stringent; however, it can constrain the $O_{bW}$ but not the $O_{bB}$.

If no deviation is observed in the $H\gamma$ production, we can set an upper limit on the Wilson coefficients at the $2\sigma$ confidence level in terms of 
\begin{equation}
\sqrt{-2 \left(n_b \ln \frac{n_s + n_b}{n_b}-n_s \right)} = 2.0,
\label{eq:significance}
\end{equation}
which yields bounds on the Wilson coefficients at the $95\%$ confidence level as follows:
\begin{align}
&\Big|c_W C_{sB} - s_W C_{sW}\Big|\left(\frac{1~ \text{TeV}}{\Lambda}\right)^2\leq 0.32,\nonumber\\ 
&\Big|c_W C_{cB} + s_W C_{cW}\Big|\left(\frac{1~ \text{TeV}}{\Lambda}\right)^2\leq 0.39,\nonumber\\ 
&\Big|c_W C_{bB} - s_W C_{bW}\Big|\left(\frac{1~ \text{TeV}}{\Lambda}\right)^2\leq 0.78;
\end{align}
see the gray bands in Fig.~\ref{cqw_cqb}. The slope of the gray bands is $-\cot\theta_W$ for the up-type quarks and $+\cot\theta_W$ for the down-type quarks, respectively. The gray bands are perpendicular to the yellow bands owing to the mixing of the weak and hypercharge fields; see Eq.~\ref{eq:MM}.

\noindent{\bf 3. The rare decay of $H \to \tau^+\tau^-\gamma$.}
\label{LeptonChannel}

\begin{table}
  \caption{Decay width and branching ratio of $H \to \tau^+ \tau^- \gamma$  with the choice of  $C_{\tau B} = 1$ or $C_{\tau W}=1$ for $\Lambda = 1~{\rm TeV} $ after demanding $ p_{\text{T}}^\gamma \geq 10~{\rm GeV}$. The SM contribution, the pure NP contribution (square) and the interference between the SM and NP effects are listed separately. }
  \label{HiggsToffGammaDecay}
  \begin{tabular}{c|c|c|c}
  \hline
   Operators & Processes & Width (GeV) & BR\\
  \hline
\multirow{2}{*}{$O_{\tau B}$}
  & Interference & $2.51 \times 10^{-6}$ & $6.27 \times 10^{-4}$ \cr\cline{2-4}
  & Square & $9.71 \times 10^{-7}$ & $2.43 \times 10^{-4}$ \cr\cline{2-4}\hline
      \multirow{2}{*}{$O_{\tau W}$}
  & Interference & $-1.35 \times 10^{-6}$ & $-3.38 \times 10^{-4}$ \cr\cline{2-4}
  & Square & $2.81 \times 10^{-7}$ & $7.04 \times 10^{-5}$ \cr\cline{2-4}\hline
      \multirow{1}{*}{SM}
  & $H\to \tau^\pm\tau^{\mp *}\to \tau^+\tau^-\gamma$ & $8.68 \times 10^{-6}$ & $2.17 \times 10^{-3}$ \cr\cline{2-4}\hline
  \end{tabular}
\end{table}

In the section we explore the HL-LHC potential of searching for the $O_{\tau B}$ and $O_{\tau W}$ through the rare decay of $H \to \tau^+\tau^-\gamma$ in the single Higgs-boson production process of $gg\to H$. 

The partial decay width of $H\to \tau^+\tau^-\gamma$ in the SM is very tiny, $\sim 8.68\times 10^{-6}~\rm{GeV}$~\cite{Kovalchuk:2017wcf} in comparison with the full width of the Higgs boson, $\sim 4~{\rm MeV}$~\cite{Heinemeyer:2013tqa}. Table~\ref{HiggsToffGammaDecay} shows the partial decay width and the branching ratio of the rare decay of $H\to \tau^+\tau^-\gamma$ induced by the operators $O_{\tau B}$ and $O_{\tau W}$ after the cut of $p_{\text T}^\gamma\geq 10~{\rm GeV}$, which is used to trigger the signal event and suppress the SM backgrounds. To be more specific, we present the interference effect and the pure NP contribution (square) separately. The NP operator effects are comparable to the SM contribution.

\begin{figure}
\includegraphics[scale=0.75]{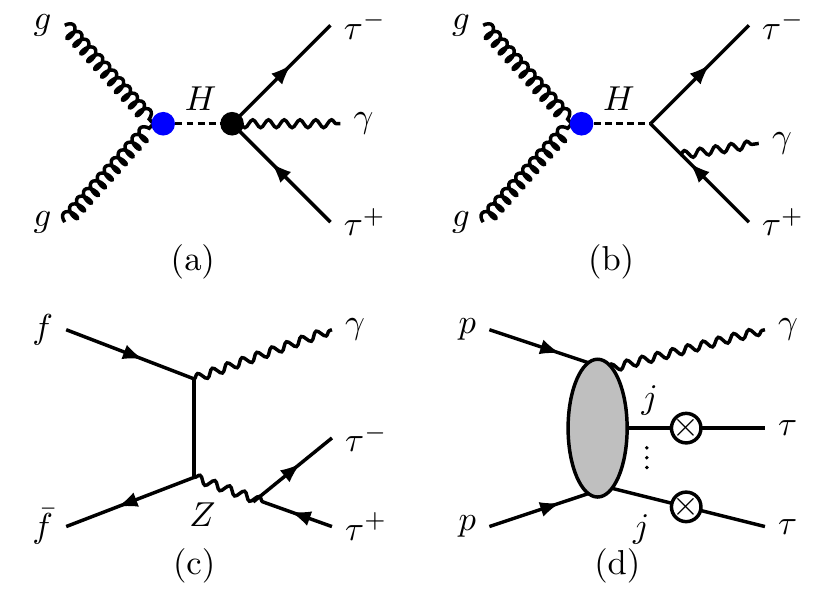}
\caption{Feynman diagrams of the signal process of $gg\to H \to \tau^+ \tau^- \gamma$ (a) and the representative diagrams of the SM backgrounds (b, c, d). } 
\label{fig:FeynmanDiagramTau}
\end{figure}

In the collider simulation we demand the $\tau^\pm$-leptons decaying hadronically; therefore, 
the signal process of $pp \to H \to \tau^+\tau^-\gamma$ yields a collider signature of two $\tau$-jets plus a hard photon. Figure~\ref{fig:FeynmanDiagramTau} displays the representative Feynman diagrams for the signal and background processes. The irreduciable backgrounds in the SM are 
\begin{eqnarray}
&& pp\to Z\gamma\to\tau^+\tau^-\gamma,\nonumber\\
&&pp\to H\to\tau^+\tau^-\gamma,
\end{eqnarray} 
and the reducible QCD backgrounds are
\begin{equation}
pp\to \gamma+\text{jets},
\end{equation}
when at least two of the QCD jets are mistagged as $\tau$-jets. 
The jet in the event is reconstructed with anti-$k_{\text{T}}$ jet algorithm~\cite{Cacciari:2008gp} with $R=0.4$. The $\tau$-tagging efficiency of the hadronic decay $\tau^\pm$ is chosen to be 60\% with the mis-tagging rate $\epsilon(j\to\tau)=1\%$. 

\begin{figure}
\centering
\includegraphics[scale=.65]{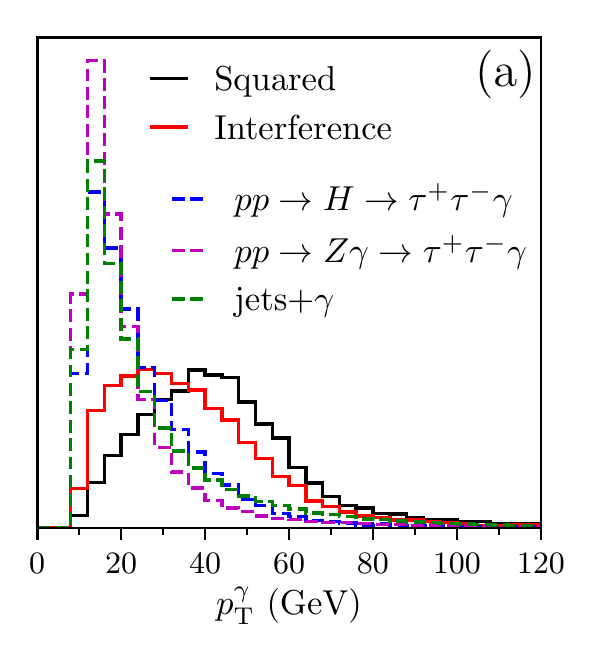}
\includegraphics[scale=.65]{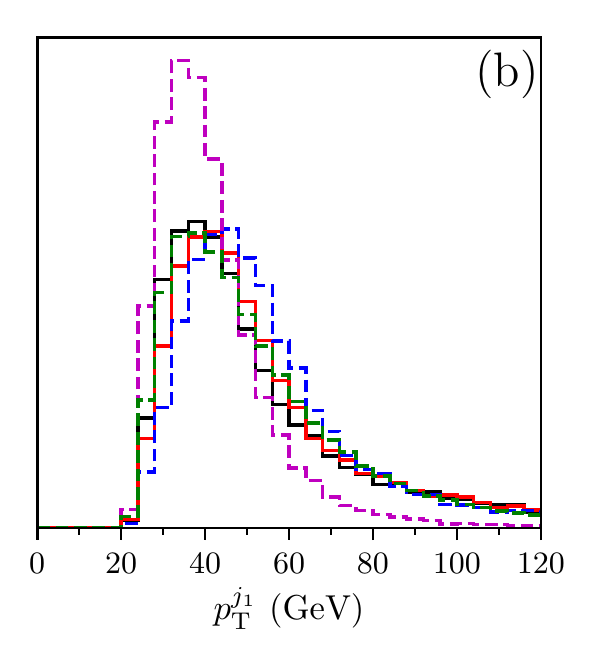}
\includegraphics[scale=.65]{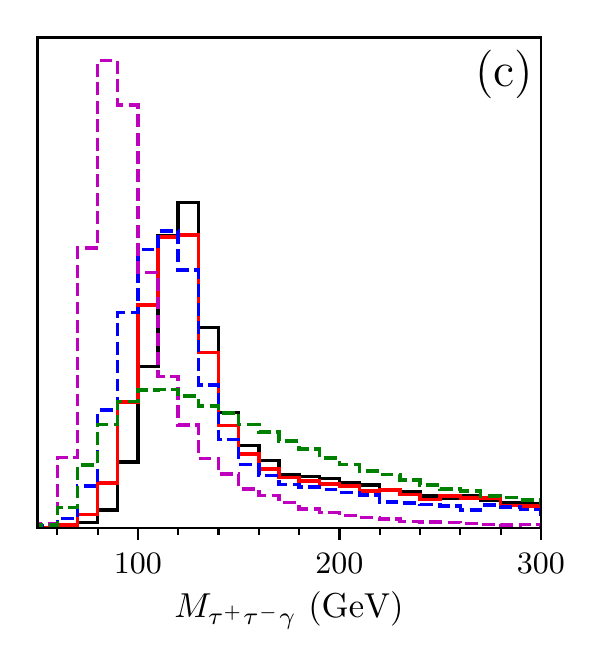}
\caption{Normalized distributions of the $p_{\text{T}}^\gamma$ (a), the $p_{\text T}^{j1}$ (b) and the invariant mass $M_{\tau^+ \tau^- \gamma}$ (c) after the basic cuts given in Eq.~\ref{eq:basic}. The black-solid and red-solid curve denotes the pure NP contribution (squared) and the interference effect (interference) in the signal event, respectively. The blue-dashed, magenta-dashed and the green-dashed curves represent the SM background processes.}
\label{DistributionLep}
\end{figure}

To tigger the signal event, we demand a set of {\it basic cuts} as follows:
\begin{align}
&p_{\text T}^\gamma\geq 10~{\rm GeV}, && |\eta_{\gamma}|\leq 2.5, \nn\\
&p_{\text T}^j\geq 20~{\rm GeV}, &&  |\eta_{j}|\leq 2.5,\nn\\
& \Delta R_{\gamma j/ jj}\geq 0.4~.
\label{eq:basic}
\end{align}
Denote $j_1$ as the $\tau$-jet with a larger $p_\text{T}$. Figure~\ref{DistributionLep} displays the normalized distributions of $p_{\text T}^\gamma$ (a), $p_{\text T}^{j_1}$ (b) and $M_{\tau^+ \tau^- \gamma}$ (c). In the signal event the photon exhibits a $p_{\text T}$ distribution harder than the $\tau$-jet. 
On the other hand, the photons in the background processes tend to be soft as they arise predominantly from the radiation of the charged leptons. In order to further suppress the SM backgrounds, we demand {\it hard cuts} on the $p_{\text T}$ of the photons and $\tau$-jets as follows:
\begin{align}
&p_{\text T}^\gamma\geq 30~\mathrm{GeV}, ~~p_{\text T}^j\geq 30~\mathrm{GeV}.
\end{align}

Figure~\ref{DistributionLep}(c) shows the normalized distributions of the invariant mass $M_{\tau^+ \tau^- \gamma}$ in which the signal process peaks around $m_H$ and one of the background  process of $pp \to Z \gamma$ peaks around $m_Z$. The $\tau$-jets in the QCD background mainly arise from the faked $\tau$-tagging and do not exhibit any resonance effect. 
Therefore, we require 
\begin{equation}
\left|M_{\tau^+ \tau^- \gamma}-m_H\right| \leq 15~\mathrm{GeV}
\end{equation}
to suppress the SM background from the process of $pp\to Z\gamma$.

\begin{table}[t]
\caption{The numbers of the signal and background events after the hard cuts and the $M_{\tau^+\tau^-\gamma}$ mass window cut at the HL-LHC for $\Lambda = 1~{\rm TeV}$. }
\label{LepEfficiency}
\begin{tabular}{c|c|c|c}
\hline
Operators & Process    & Hard cuts & $M_{\tau^+ \tau^- \gamma}$  \\
\hline
\multirow{2}{*}{ $C_{\tau B}=1$, $C_{\tau W}=0$}
& Interference  & $1.16\times 10^{3}$ & $331$\cr\cline{2-4}
& Square   & $470$ & $139$\cr\cline{2-4}\hline
 \multirow{2}{*}{ $C_{\tau B}=0$,~$C_{\tau W}=1$}
& Interference   & $-624$ & $-178$\cr\cline{2-4}
&Square   & $136$ & $40.2$\cr\cline{2-4}\hline\hline
\multirow{3}{*}{Backgrounds}
& $pp\to H \to \tau^+ \tau^- \gamma $& $5.75\times 10^{3}$ & $1.63\times 10^{3}$\cr\cline{2-4}
&$pp\to Z\gamma\to \tau^+ \tau^- \gamma$  & $4.40\times 10^{4}$ & $1.01 \times 10^{4}$\cr\cline{2-4}
&$pp\to \gamma + $jets   & $8.11 \times 10^{6}$ & $1.85 \times 10^{5}$\cr\cline{2-4}\hline
\end{tabular}
\end{table}

Table~\ref{LepEfficiency} shows the numbers of the signal and the background events at the HL-LHC after the hard cuts and the mass window cut of $M_{\tau^+\tau^-\gamma}$. We separate the signal contribution into the pure NP effect (square) and the interference effect. 
Again, the $O_{\tau B}$ and $O_{\tau W}$ yield exactly the same cut efficiencies. As a result, the number of the signal events for a general choice of $C_{\tau B}$ and $C_{\tau W}$ can be expressed as follows: 
\begin{eqnarray}
n_s &=& n_s^{\text{sqr}}\times\frac{(c_W C_{\tau B} - s_W C_{\tau W})^2}{c_W^2} \left(\frac{1~ \text{TeV}}{\Lambda}\right)^4 \nonumber\\
&+&n_s^{\text{int}}\times\frac{(c_W C_{\tau B} - s_W C_{\tau W})}{c_W} \left(\frac{1~ \text{TeV}}{\Lambda}\right)^2,
\end{eqnarray}
where $n_s^{\text{sqr}}$ is the signal event number from the pure NP contribution (square) and  $n_s^{\text{int}}$ is the signal event number from the interference effect (interference) for $C_{\tau B}=1$ and $C_{\tau W}=0$. 

\begin{figure}
\centering
\includegraphics[scale=.55]{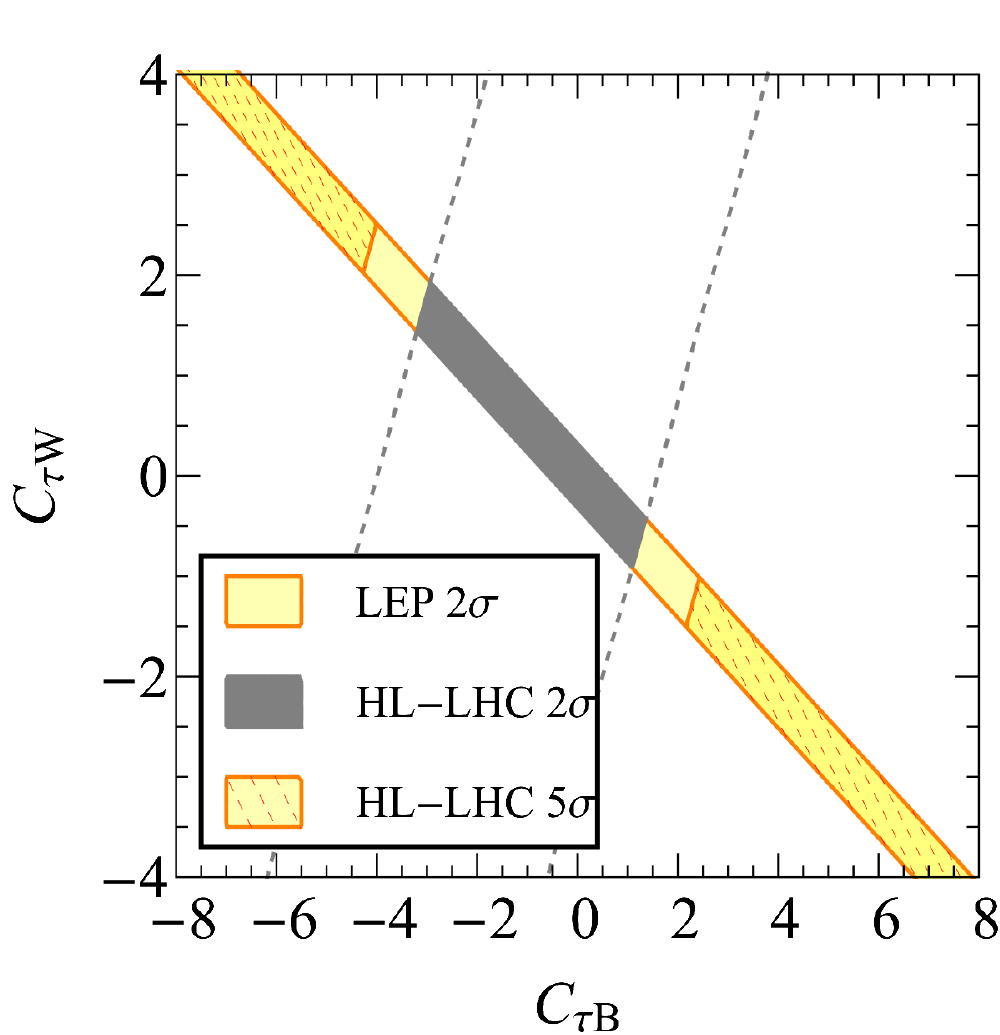}
\caption{The yellow-meshed bands denote the $5\sigma$ discovery region of $C_{\tau B}$ and $C_{\tau W}$ in the the process of $gg\to H\to \tau^+\tau^-\gamma$ at the HL-LHC with $\Lambda=1~{\rm TeV}$.  The yellow bands denote the $2\sigma$ parameter space allowed by the $Z$-pole measurement at the LEP. The region between the two dashed lines is allowed at the $2\sigma$ significance if no NP effects are observed at the HL-LHC, where the overlapped gray band satisfies both the LEP and the HL-LHC bounds.}
\label{clw_clb}
\end{figure}

Using Eq.~\ref{eq:dis},  we obtain that a $5\sigma$ discovery significance in the process of $gg\to H\to \tau^+\tau^-\gamma$ requires
\begin{align}
\Big(c_W C_{\tau B} - s_W C_{\tau W}\Big)\left(\frac{1~ \text{TeV}}{\Lambda}\right)^2 \leq -4.7,
\end{align}
or
\begin{align}
\Big(c_W C_{\tau B} - s_W C_{\tau W}\Big)\left(\frac{1~ \text{TeV}}{\Lambda}\right)^2 \geq 2.6~.
\end{align}
Figure~\ref{clw_clb} shows the parameter space of $5\sigma$ discovery in the plane of $C_{\tau B}$ and $C_{\tau W}$ obtained from the process of $gg\to H\to \tau^+\tau^-\gamma$ at the HL-LHC (yellow-meshed band) with $\Lambda=1~{\rm TeV}$. The yellow band denotes the $2\sigma$ parameter space allowed by the LEP measurement~\cite{Escribano:1993xr}. The HL-LHC can cover the most of the parameter space that cannot be accessed at the LEP. 

If no deviation is found in the process of $gg\to H\to \tau^+\tau^-\gamma$, we obtain a bound on the Wilson coefficients at the $95\%$ confidence level as 
\begin{equation}
-3.5 \leq \Big(c_W C_{\tau B} - s_W C_{\tau W}\Big)\left(\frac{1~ \text{TeV}}{\Lambda}\right)^2 \leq 1.4;
\end{equation}
see the region between the two dashed lines. The overlapped gray band satisfies both the LEP and the HL-LHC bounds.

~\\
\noindent{\bf 4. Conclusion.}
\label{section5}

The magnetic moment and weak magnetic moment of the SM fermions are sensitive to quantum effects of new physics resonances. For each fermion $f$ there are two independent operators that generate the magnetic moment and weak magnetic moment; one is $O_{fB}$ involving the hypercharge field, the other is $O_{fW}$ involving the weak field. After symmetry breaking the magnetic moment and the weak magnetic moment depend on the orthogonal combinations of the two operators. Therefore, at least two independent experiments are needed to probe the $O_{fB}$ and $O_{fW}$. 

The weak magnetic moment of the strange-quark ($s$), charm-quark ($c$) or  bottom-quark ($b$) is bounded by the width measurement of the $Z$-boson at the LEP, but the magnetic moments of the three quarks are less constrained. In this paper, we explore the potential of the HL-LHC on probing the operators $O_{f B}$ and $O_{f W}$ ($f=s,c,b$) in the process of $pp\to H\gamma$, in which the magnetic moment of the quarks dominates. We consider one flavor of quarks at a time. We showed that, in most of the parameter space that cannot be accessed at the LEP, a $5\sigma$ significance discovery can be reached in the $H\gamma$ production at the HL-LHC.

The magnetic moment and weak magnetic moment of the electron or muon have been accurately measured and thus severely constrained. In analogue to the electron and muon leptons the weak magnetic moment of the $\tau$ lepton is also bounded at the LEP; however, the magnetic moment of the $\tau$ lepton is less bounded. In this work we consider the rare decay of $H\to \tau^+\tau^-\gamma$ induced by the $O_{\tau B}$ and $O_{\tau W}$ and examined the potential of probing the two operators in the process of $gg\to H \to \tau^+\tau^-\gamma$. Similar to the case of quark operators, in most of the parameter space that cannot be accessed at the LEP, a $5\sigma$ significance discovery can be reached in the process of $gg\to H\to \tau^+\tau^-\gamma$ at the HL-LHC. 

In summary, one can probe the magnetic moment of the $s(c,b)$ quarks in the process of $pp\to H\gamma$ and the magnetic moment of the $\tau^\pm$ lepton in the $gg\to H\to\tau^+\tau^-\gamma$ at the HL-LHC.  

~\\
\noindent{\bf Acknowledgement.}

The work is supported in part by the National Science Foundation of China under Grant Nos. 11725520, 11675002, 11635001, 11805013, 12075257 and the Fundamental Research Funds for the Central Universities under Grant No. 2018NTST09. 
	
\bibliography{draft}
\bibliographystyle{apsrev}
	
\end{document}